# Molecular Model Checking a Temporal Logic

Weijun ZHU

School of Information Engineering, Zhengzhou University, Zhengzhou, 450001, China

E-mail: zhuweijun@zzu.edu.cn

**Abstract.** The molecular computing has been successfully employed to solve more and more complex computation problems. However, as an important complex problem, the model checking problem are still far from fully resolved under the circumstance of molecular computing, since it is still a lack of method. To address this issue, a model checking method is presented for checking the basic constructs in a given temporal logic using molecular computing. Through the design of the new encoding and calling this process, we can get a molecule-based approach for checking all of the basic constructs of this logic.

## I Introduction

Differ from an electronic computer, a DNA computer use DNA molecules as the carrier of computation. In 1994, a Turing Award winner professor Adleman published an article in <Science> that solved a small scale Hamilton path problem with a DNA experiment [8], and it is considered the pioneer work in the field of DNA computing. Since the DNA computing has a huge advantage of parallel computing, this technique was subsequently developed rapidly. After this famous experiment, many models and methods based on DNA computing were presented for solving some complex computational problems, especially the famous NP-hard problems and PSPACE-hard ones. For examples, Lipton published an article in <Science> that promoted Adleman's idea and tried to solve the SAT problem [9], Ouyang et al. published an article in <Science> that presented a DNA-computing-based model for solving the maximal clique problem [10], Shapiro published an article in <Nature> that solved an automata problem of two states and two characters using the autonomous DNA computing technique [11].

On the one hand, the technique based on biochemical reactions in test tubes, the one based on nano devices and the one based on molecular self-assembly can be applied to solve some problems in computer science [8][12][13][14]. On the other hand, due to the excellent information processing mechanism and the huge parallelism, some living cells can also be employed to perform some computations. The site-specific DNA recombinase Hin, which can mediate inversion of DNA segments that represent variables, was used to produce the solution. In this model, each cell can produce and examine a solution of satisfiablity problem. As a result, billions of cells can explore billions of possible solutions [15]. In this way, Prof. Xu constructed a cellular computing model in [15] which can solve satisfiablity problem.

One of the key differences between computer and other computing tools is the universality. Prof. Xu constructed a mathematical model called "probe machine" for the general DNA computer [17]. By integrating the storage system, operation system, detection system and control system into a whole, he gradually obtains a real general DNA computer --- "Zhongzhou DNA computer" [17]. According to Ref.[16], a probe machine is a nine-tuples consisting of data library, probe library, data controller, probe controller, probe operation, computing platform, detector, true solution storage and residue collector. It is an universal DNA computing model which can be realized in biology, and a Turing machine is just a "special case" of a probe machine [16]. This significant progress has raised the practical importance of the researches on DNA computing.

Beside the satisfiablity problem, the Model Checking (MC) one is another important computational problem. And the two problems are correlative. The MC was proposed by the Turing Award winner Prof. Clarke et al [1]. The MC algorithms answer automatically the question whether a system satisfies the given property or not. The model checking is widely used in the fields of CPU verification [2], network protocol verification, security protocol verification [3], software verification [4]. NASA, Intel, IBM and Motorola are using this technique. The general principles of MC can be given as follows. A system model is constructed with an automaton, a property which the system should satisfies is described by a temporal logic formula. If the automaton is a model of the formula, then the system model satisfies the properties, otherwise, the system doesn't satisfy the properties.

In order to describe the different temporal properties, the researchers have presented some different temporal logics. The Turing Award winner Prof. Pnueli introduced Linear Temporal Logic (LTL for short) into computer science in [5], and this logic can express the linear properties. The Turing Award winner Prof. Clarke proposed Computation Tree Logic (CTL for short) in [6][7], and this logic can express the branch properties.

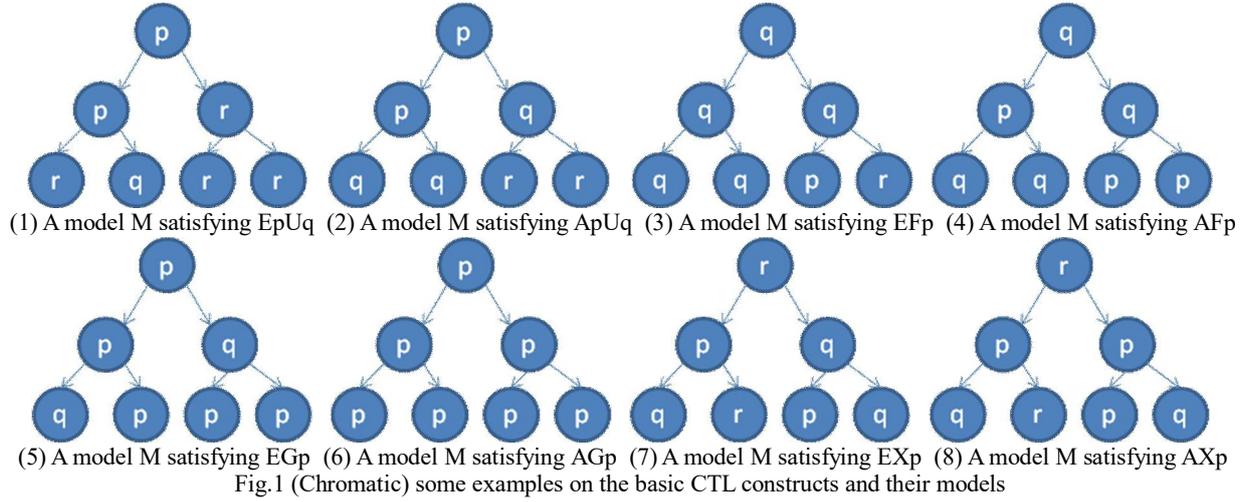

(1) A model M satisfying EpUq  (2) A model M satisfying ApUq  (3) A model M satisfying EFp  (4) A model M satisfying AFp

(5) A model M satisfying EGp  (6) A model M satisfying AGp  (7) A model M satisfying EXp  (8) A model M satisfying AXp

Fig.1 (Chromatic) some examples on the basic CTL constructs and their models

As a complex computational problem, the model checking under the circumstance of DNA computing is always the goal of researchers. In 2006, the Turing Award winner Prof. Emerson employed some DNA molecules to conduct CTL model checking for the first time [18]. As for LTL, the model checking is a PSPACE problem in the classical computing, and we have found a DNA-computing-based method, see Ref.[19] and Ref.[20], which can be used for checking all four basic constructs and some popular formulas. The basic constructs in CTL are: EpUq, ApUq, EFp, AFp, EGp, AGp, EXp, AXp. We can obtain arbitrary CTL formula by combining these basic constructs recursively. Up to now, many basic constructs in CTL cannot be conducted model checking within the framework of DNA computing. This is the problem to be solved in this paper.

## II preliminary

### 2.1 The basic constructs in CTL [1]

**Definition 1** Let p and q be atomic propositions, EpUq, ApUq, EFp, AFp, EGp, AGp, EXp and AXp be the basic CTL construct. An arbitrary CTL formula can be obtained by combining recursively some basic CTL constructs. An atomic proposition and a basic CTL construct are interpreted on a system model M, and their intuitive meanings are given as follows.

● p or q is satisfied in a state s, or not.

● EpUq describes the following property: There exists at least one path in M, such that p is always satisfied until q is satisfied.

● ApUq describes the following property: For each path in M, p is always satisfied until q is satisfied.

● EFp describes the following property: There exists at least one path in M, such that p is eventually satisfied.

● AFp describes the following property: For each path in M, p is eventually satisfied.

● EGp describes the following property: There exists at least one path in M, such that p is always satisfied.

● AGp describes the following property: For each path in M, p is always satisfied.

● EXp describes the following property: There exists at least one path in M, such that p is satisfied in the next state.

● AXp describes the following property: For each path in M, p is satisfied in the next state.

Fig.1 gives an example for each basic CTL construct, respectively.

Given an arbitrary model M, how to use the DNA-computing-based method to determine whether the basic CTL constructs be satisfied by M or not? To this end, Section 3.1 will give such an approach.

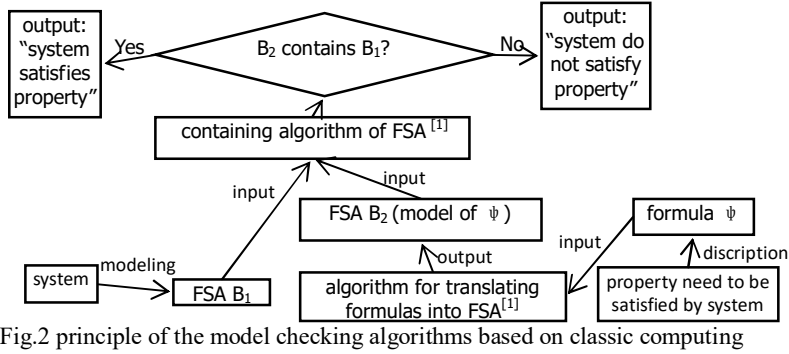
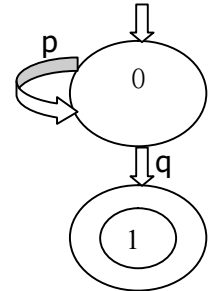

Fig.2 principle of the model checking algorithms based on classic computing    Fig.3 an example on FSA

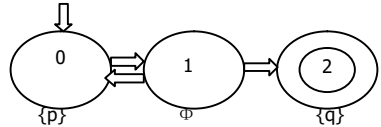

Fig.4 an example on LFSA: the systematic FSA model $M_1$ which is used in the experiments in this paper

## 2.2 Finite state automata and model checking

**Definition 2** A Finite State Automaton (FSA) is a five-tuples ($\Sigma$,Q,T,q0,F), where

● $\Sigma$ is a finite alphabet

● Q is a finite set of states

● T is a finite set of transitions: $T: Q \times \Sigma \to R(Q)$

● $q_0 \in Q$ is an initial state

● $F \subseteq Q$ is a set of acceptance states

Fig.3 gives an example for a FSA. This automaton is made up of the two states and the two transitions. The state 0 is an initial state which is pointed by an arrow without source, whereas the state 1 is an acceptance state which is marked by a double circle. The automaton will enter the state 0 if p is input at the state 0, whereas the automaton will enter the state 1 if q is input at the state 0. The string pq is an acceptance word, since the automaton will transit from an initial state to an acceptance state if pq is input. Similarly, the strings q, ppq, pppq, ... are acceptance words too. An acceptance language of an automaton is made up of all of the acceptance words of the automaton. In this example, {q、pq、ppq、pppq...} is the acceptance language of the automaton which is illustrated by Fig.3.

The only difference between the automaton in Fig.3 and the one in Fig.4 is that the atomic propositions in the latter automaton are satisfied in the states rather than in the transitions. Therefore, the latter automaton is called a Label FSA (LFSA).

In classical computation, the principles of the algorithms for temporal logic model checking can be illustrated by Fig.2. A LFSA, denoted as $B_1$, is used to describe some behaviors of a system, whereas a FSA, denoted as $B_2$, is employed to construct a model of a temporal logic formula. The model checking algorithm will decide that the system meets the property specified by the formula, if some inclusion relations hold between the two acceptance language of the two automata.

## 2.3 Sticker automata and DNA model checking

### 2.3.1 Sticker automata

As a model of DNA computing, a sticker automaton can realize a FSA. Given a DNA strand charactering an input string and a FSA, the sticker automaton can determine whether or not the string is accepted by the FSA.

2.3.1.1 The encoding way of FSA and input string

Ref.[21] gives the following way of DNA encoding

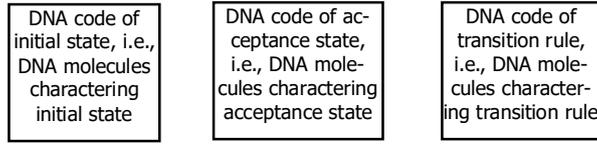

(1) DNA molecules charactering run of systematic FSA, i.e., the class I molecules)

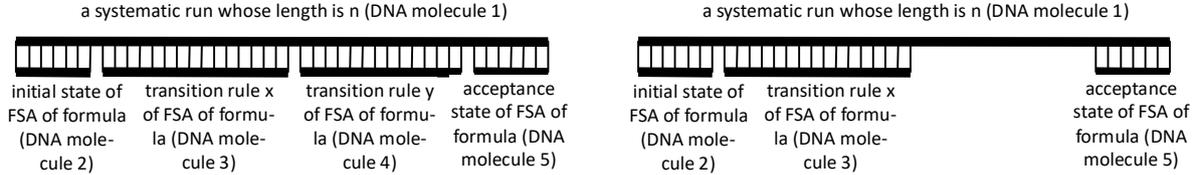

(2) three kinds of DNA molecules charactering FSA of formula, i.e., the class II molecules

(3) A complete double strand is formed after hybridization  (4) An incomplete double strand is formed after hybridization
if a run is accepted by the FSA model of formula        if a run cannot be accepted by the FSA model of formula
where DNA molecule 1 is a class I molecule, and DNA molecule 2, 3, 4, 5 are class II molecules

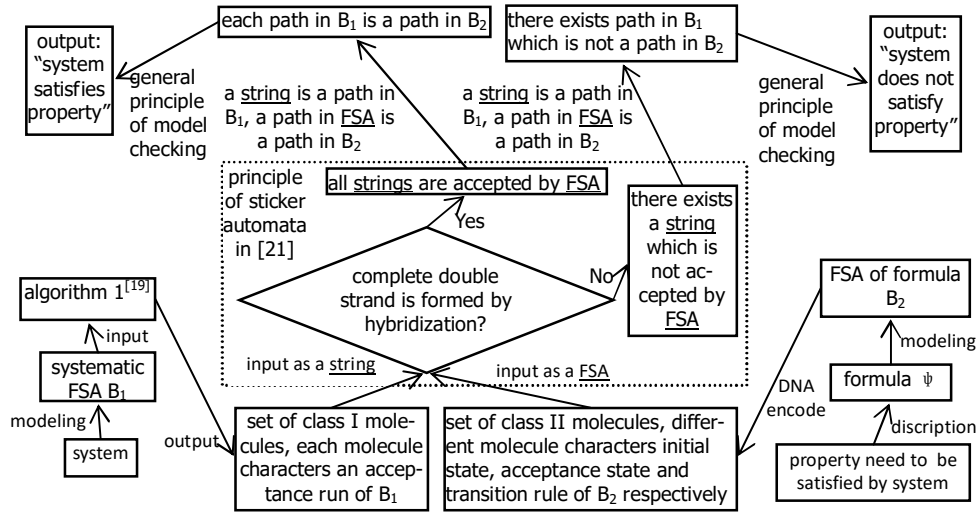

(5) Flowchart of the algorithms
Fig.5 principle of the LTL model checking algorithms based on DNA computing [20]

Supposing M=($\Sigma$,S,T,$s_0$,F) is a FSA, and every character a in the alphabet $\Sigma$ can be encoded as C(a), we have:

(1) An input string $a_1,...,a_n$ in $\Sigma$ can be encoded with the following single-stranded DNA molecule: 5' $I_1$ $X_0...X_m$ $C(a_1)$ ... $X_0...X_m$ $C(a_n)$ $X_0...X_m$ $I_2$ 3', where $I_1$ is an initiator sequence, $X_0...X_m$ is a spacer sequence separating $C(a_i)$, and $I_1$ is a terminator sequence.

(2) A transition T($s_i$,a)=$s_j$ is encoded as 3' $\overline{X_{i+1}...X_m}$ $\overline{C(a)}$ $\overline{X_0...X_j}$ 5', where $\overline{X}$ means the Watson-Crick complement (WC for short) of a nucleotide X, $\overline{C(a)}$ means the WC of the DNA strand charactering a.

(3) An initial state $s_i$ is encoded as 3' $\overline{I_1}$ $\overline{X_0...X_i}$ 5'.

(4) An acceptance state $s_j$ is encoded as 3' $\overline{X_{j+1}...X_m}$ $\overline{I_2}$ 5'.

2.3.1.2 The process of DNA computing based on sticker automata

The computational process of sticker automata can be concluded as follows [21].

Step 1: data preprocessing

(1) Synthesize some DNA strands charactering an automaton and its input strings.

(2) Put all the DNA strands into the test tube T, and anneal to make sure that the strands and their WC complements

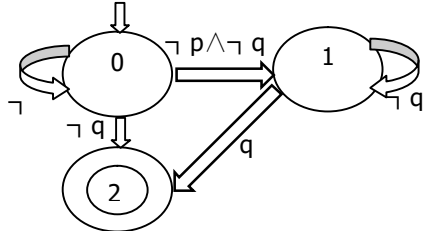

Fig.6 a FSA model $A_1$ of the one formula

Tab.1 relationships: a formula and its FSA model

| The formula | $\varphi_1 = \neg p \overline{U} \neg q$ |
|---|---|
| FSA of formula | $A_1$ |

can be hybridized completely. The process of base pairing and the placement of ligase can form complete or partial double stranded DNA molecules.

Step 2: computation

After the first step, we will see the following phenomena. If the input string is accepted by the automaton, the tube T contains only the complete double stranded DNA molecules which begin with an initiator sequence and terminate at a terminator sequence. Otherwise, there are partial double stranded DNA molecules or single stranded DNA molecules in T. The main reason for the latter case is that, some fragments of the single stranded DNA molecules charactering input strings are paired successfully with some single stranded DNA molecules charactering transitions, whereas other fragments of the single stranded DNA molecules charactering input strings cannot be paired with any single stranded DNA molecules charactering transitions. Therefore, we add some ribozymes called Mung Bean into the test tube T, in order to degrade the single stranded DNA fragment, and retain the complete double stranded DNA molecules.

Step 3: output of results

We can separate the different DNA molecules with different lengths using electrophoretic technique. If there exists a variety of length of DNA molecules, it indicates that there are some partial double stranded DNA molecules in T before we add the ribozymes, and the input string cannot be accepted by the automaton. Otherwise, T contains only complete double stranded DNA molecules before we add the ribozymes, and the input string can be accepted by the automaton.

### 2.3.2 DNA model checking

On the basis of sticker automata, Ref.[20] prenseted a DNA-computing-based LTL model checking method, which can be denoted as the algorithm TL-MC-DNA(DNACODE(A),x), where DNACODE(A) and x are two input of the algorithm, A is a FSA expressing a run of a system, DNACODE(A) is an encoding with a sticker automaton for charactering A, x=DNACODE(A(f)) is an encoding with a sticker automaton for charactering A(f), and A(f) is a FSA model of a formula f. In Ref.[20], the scope of f includes the all the basic LTL formulas and some popular LTL formulas, f formula for short. The output of the algorithm is yes or no, reprenting the result of the model checking. The principle of this algorithm is illustrated by Fig.5.

### 2.4 A FSA of the formula and its DNA model checking

Given a temporal logic formula, one can compute its FSA model [1]. Fig.6 gives a FSA model for one specific formula. The corresponding relations are shown in Tab.1. It should be noted that, $\neg$ is logical negation, $\overline{U}$ is logical duality of U, and $\neg p \overline{U} \neg q$ describes the following property: For each path in M, there exists at least one state which does not satisfy p, before q is satisfied.

## III The DNA model checking method for the basic CTL constructs

As mentioned in section 2.3.2, if the encoding of sticker automaton which realizes a FSA of a system and the encoding of sticker automaton which realizes a FSA of a formula are inputted, the algorithm TL-MC-DNA(DNACODE(A),DNACODE(A(f))) in [20] can compute and return a result of model checking. This paper expands range of f, and a seires of new encoding of sticker automata, see section 4. By computing TL-MC-DNA(DNACODE(A), DNACODE(A(f''))), where f''={$\varphi_1$}, (f'' formula for short), we can perform DNA model checking for the one temporal logic formula in Tab.1. Ref.[20] has confirmed the effectiveness of the algorithm TL-MC-DNA for the f formulas by simulated biological experiments. Section 4 will study the effectiveness of the new algorithm for the f'' formulas by a number of simulated biological experiments.

It should be noted that the algorithm TL-MC-DNA(DNACODE(A),DNACODE(A(f))) comes from Ref.[20], and its pseudo code is no longer given, due to limitations on space.

### 3.1 The DNA model checking for the four universal formulas

ApUq, AFp, AGp and AXp are called the universal formulas since their semantics are all involved in "all paths".

Comparing the CTL formula ApUq and the LTL formula pUq, we can clearly see that these two formulas have the same semantics. Therefore, we can use the algorithm TL-MC-DNA (DNACODE (A), x) to check the CTL formula ApUq. Similarly, the algorithm TL-MC-DNA (DNACODE (A), x) can also be employed to check the CTL constructs AFp, AGp and AXp. The obtained algorithm is formulated as follows.

---

**ALgorithm 1**. The DNA model checking algorithm for the universal CTL formulas CTLQ-MC-DNA(DNACODE(A), DNACODE(A($f_q$))).
**INPUT**: the encoding of sticker automaton which realizes a systematic FSA A, the encoding of sticker automaton which realizes a FSA of an universal CTL formula $f_q$, where $f_q$=ApUq, AFp, AGp or AXp.
**OUTPUT**: whether A satisfies $f_q$, or not
**BEGIN**
Step 1:
SELECT CASE $f_q$
  CASE ApUq
    g:=pUq  //where g is a f formula
  CASE AFp
    g:=Fp  // where g is a f formula
  CASE AGp
    g:=Gp  // where g is a f formula
  CASE AXp
    g:=Xp  // where g is a f formula
ENDSELECT
Step 2: y:=TL-MC-DNA(DNACODE (A), DNACODE (A(g))
Step 3: IF y="yes",THEN return "yes", ELSE return "no"
**END**

---

### 3.2 The DNA model checking for the four existence formulas

EpUq, EFp, EGp and EXp are called the existence formulas since their semantics are all involved in "there exists at least one path".

The formula EpUq and the formula ApUq have the following relationship: ¬ EpUq=A¬ pŪ¬ q. That is to say, EpUq=¬ A¬ pŪ¬ q, where A¬ pŪ¬ q describe the following property: For each path in M, there exists at least one state which does not satisfy p, before q is satisfied.

The formula EGp and the formula AFp have the following relationship: ¬ EGp=AF¬ p, that is to say, EGp=¬ AF ¬ p.

The formula EFp and the formula AGp have the following relationship: ¬ EFp=AG¬ p, that is to say, EFp=¬ AG ¬ p.

The formula EXp and the formula AXp have the following relationship: ¬ EXp=AX¬ p, that is to say, EXp=¬ AX ¬ p.

Comparing $A\neg pU\neg q$ and $\varphi_1=\neg pU\neg q$, we can clearly see that these two formulas have the same semantics. Thus, $\neg \varphi_1=EpUq$. Therefore, we can use the algorithm TL-MC-DNA (DNACODE (A), DNACODE (A(f''=$\varphi_1$))) to check the CTL formula EpUq. Similarly, the algorithm TL-MC-DNA (DNACODE (A), x) can also be employed to check the CTL formulas EFp, EGp and EXp. The obtained algorithm is formulated as follows. It should be noted that, only one new atomic proposition rather than any modifications to the algorithm, FSA structure or sticker automata encoding scheme, is needed in the design of DNA encoding, when a negative form of an atomic proposition occurs in the algorithm and be as its argument.

**Algorithm 2**. The DNA model checking algorithm for the existence CTL formulas CTLC-MC-DNA(DNACODE(A), DNACODE(A($f_c$))).
**INPUT:** the encoding of sticker automaton which realizes a systematic FSA A, the encoding of sticker automaton which realizes a FSA of an existence CTL formula $f_c$, where $f_c$=EpUq, EFp, EGp or EXp
**OUTPUT**: whether A satisfies $f_c$, or not

**BEGIN**
SELECT CASE $f_c$
  CASE EpUq
    Step 1: g:= $\varphi_1$
    Step 2: y:=TL-MC-DNA（DNACODE(A)，DNACODE（A（g）） // where g is a f'' formula
    Step 3: IF y="yes",THEN return "no", ELSE return "yes" //$\varphi_1$=$\neg$ (EpUq)
  CASE EFp
    Step 1: g:=G$\neg$ p
    Step 2: y:=TL-MC-DNA（DNACODE(A)，DNACODE（A（g）） // where g is a f formula
    Step 3: IF y="yes",THEN return "no", ELSE return "yes" // G$\neg$ p=$\neg$ (EFp)
  CASE EGp
    Step 1: g:= F$\neg$ p
    Step 2: y:=TL-MC-DNA（DNACODE(A)，DNACODE（A（g）） // where g is a f formula
    Step 3: IF y="yes",THEN return "no", ELSE return "yes" // F$\neg$ p=$\neg$ (EGp)
  CASE EXp
    Step 1: g:=X$\neg$ p
    Step 2: y:=TL-MC-DNA（DNACODE(A)，DNACODE（A（g）） // where g is a f formula
    Step 3: IF y="yes",THEN return "no", ELSE return "yes" // X$\neg$ p=$\neg$ (EXp)
ENDSELECT
**END**

### 3.3 The DNA model checking for the basic CTL constructs

The principle of this algorithm is: (1) If a basic CTL construct is an universal formula, the algorithm 1 will be called. (2) And if a basic CTL construct is an existence formula, the algorithm 2 will be called. In this way, the model checking of the basic CTL constructs can be conducted. The algorithm is formulated as follows.

**Algorithm 3**. The DNA model checking algorithm for the basic CTL constructs CTL-MC-DNA(DNACODE(A), DNACODE(A($f_{CTL}$))).
**INPUT**: the encoding of sticker automaton which realizes a systematic FSA A, the encoding of sticker automaton which realizes a FSA of a basic CTL construct $f_{CTL}$
**OUTPUT**: whether A satisfies $f_{CTL}$, or not

**BEGIN**
Step 1: IF there exists $f_c$, such that $f_{CTL}=f_c$, THEN call CTLC-MC-DNA(DNACODE(A), DNACODE(A($f_c$)))
     ELSEIF there exists $f_q$, such that $f_{CTL}=f_q$, THEN call CTLQ-MC-DNA(DNACODE(A), DNACODE(A($f_q$)))
**END**

### 3.4 Complexity analysis

The time complexity of the algorithm TL-MC-DNA is O(m+n) [20], where m means the number of nodes in an automaton, and n means the number of edges in this automaton. Therefore, the algorithm 1 needs execute O(m+n)+ O(3)= O(m+n) times operations. Similarly, the algorithm 2 needs execute O(m+n)+ O(3)= O(m+n) times operations. The algorithm 3 calls the algorithm 1 or the algorithm 2, so that the complexity of the algorithm 3 is O(m+n). In comparison, the model checking of the basic CTL constructs based on classical computing has a square complexity.

# IV Simulated experiments

As far as the new method is concerned, the only thing that is related to the realization of molecular biology is the TL-MC-DNA algorithm. With a simulation platform called NUPACK in [22], Ref.[20] has confirmed that: (1) for the nine FSAs of the nine specific temporal logic formulas, the algorithm TL-MC-DNA can be realized effectively in molecular biology; (2) for the above FSAs, one can design their appropriate encoding of sticker automata, so that the accuracy rate of base pairing reaches more than 99%. As for the FSA of formula presented in section 2.4, Can the TL-MC-DNA algorithm be implemented effectively in molecular biology? Section 4 employees the same experimental platform and experimental means with the ones in [20] to carry out some molecular biological simulated experiments. We need to examine the thermal denaturation temperature and the free energy [20].

**Experimental procedure**: (1) according to Fig.4 and Fig.6, one can design the encoding of the sticker automata for the systematic FSA, as well as the encoding of the sticker automata for the FSA of formula, respectively; (2) for these FSAs mentioned above, one can simulate the process of hybridization between some single stranded DNA molecules; (3) according to the algorithms proposed in this paper, one can get the results of model checking of the several formulas, by reading the results of hybridization.

**Experimental objective**: To test the correctness, effectiveness and biological realizability of the new algorithms.

## 4.1 Simulated experiments for $\varphi_1$

### 4.1.1 Encoding designs

We have designed a DNA encoding via NUPACK, as shown in Tab.2. Fig.7, Fig.8 and Fig.9 show the thermodynamic analysis of the encoding sequence presented in Tab.2 at 10 Celsius degree.

As shown in Fig.7, the Normalized Ensemble Defect (NED) means the incorrect matching ratio of the nucleotides when a biochemical reaction is in equilibrium. 0% tips for an optimal design, whereas 100% tips for the worst design. The NED of our coding sequence is 0.1%.

The principle of the minimum free energy points out that the free energy is minimized when a biochemical reaction is in equilibrium. As shown in Fig.8, the color of the match between two kinds of molecules is dark red. The probability of the event that double stranded molecule completely matched and it reached the balance, almost reach 100%, by comparing color change of vertical bar that indicate the balance probability. Thus, its free energy is approximately equal to the minimum free energy.

As shown in Fig.9, the position of the red line indicates that all bases in the two single strands are completely complementary to each other, and the color of the red line indicate that the probability of all the pairs are approximately equal to 1.

On the basis of the above analysis, we can say that our DNA sequence satisfies the minimum free energy constraint, and the DNA molecules that participate in the reaction have a basically consistent temperature of solution chain. Therefore, the experimental results obtained from this encoding are reliable and effective in biologic.

In fact, Tab.2 indicates the encoding rules for the input strings, as shown in Tab.3. According to Tab.3 and the principle of encoding of sticker automata, we can deduce the encoding of the sticker automaton charactering $\varphi_1$, as shown in Tab.4.

### 4.1.2 Simulated experiments

With the DNA code given in section 4.1.1 at hands, we can conduct our simulated experiments. It should be noted that, in section 4.1.2, all the encoding of the DNA molecules are written from left to right with a 5'-3' direction, which is consistent with the way of writing in NUPACK.

We will check whether or not the systematic FSA $M_1$ satisfies the formula $\varphi_1$. According to the DNA codes given by section 4.1.1, we can get all the paths which come from the systematic runs, as shown in Tab.5, where k is a

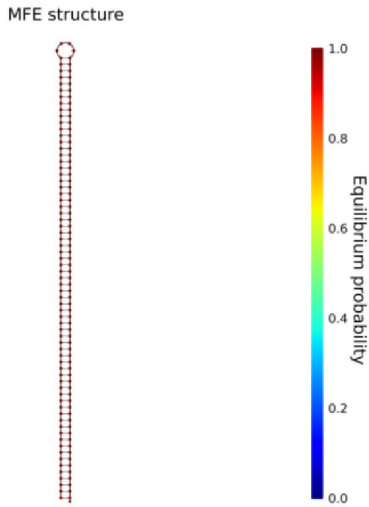
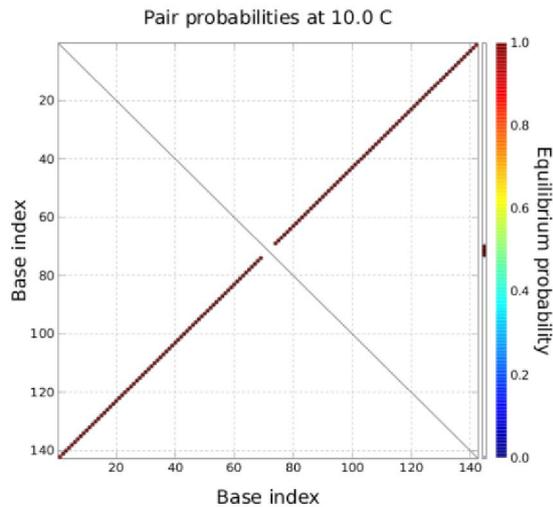

Fig.7 checking the formula $\varphi_1$: the structural properties of encoding sequence natural number. By observing the transition rules shown in Tab.4, we are fully aware that none of atomic proposition excerpt for s, u and q takes part in the transitions of states. Therefore, we do no need to consider whether or not the states satisfy the atomic propositions p and r.

Fig.8 (Chromatic) thermodynamic analysis for $\varphi_1$: minimum free energy structure

Fig.9 (Chromatic) checking for $\varphi_1$: pairing probability in equilibrium

First, we will check the path 1. There are two possible runs in this path. Without loss of generality, we support that the atomic proposition sequence which is crossed by the run is suq.

All the molecules expressing the runs begin with GCCAGAA and end with GGCCGTC. Thus, we only need to deal with d=TTGCAAGGCAGCGAATTGCAAGGCGCGGAATTGCAAGGCCCCGAATTGCAA. In short, we will observe whether or not hybridization occurs between the DNA molecules expressing transitions and the molecule d. To this end, we pour the following six kinds of molecules into a container with a volume of $10^{-15}$L: d, t0s0, t0u1, t0s2, t1s1 and t1q2, for observing the hybridization.

The systematic run which is expressed by the molecule d, crosses the three states. If hybridization occurs between the DNA molecules expressing transitions and the molecule d, there are not more than three kinds of molecules which are the WC of some segment of d, involved in the specific hybridization. For selecting three kinds of molecules from all the five kinds of WC molecules, one has ten choices. Thus, we execute the following ten groups of sub-experiments, accordingly.

 (1) Group 1: t0s0, t0u1, t1q2 and d

The concentrations of the four kinds of molecules reach respectively 100uM, 100uM, 100uM and 100uM, and their molecular numbers are respectively 60000, 60000, 60000 and 60000. With the temperature naturally dropped to 10 Celsius degree, the hybridization reaction is observed. Fig.10(1) and Fig.10(2) show the result of the hybridization, where strand1 means d, strand2 means t0s0, strand3 means t0u1 and strand4 means t1q2.

See Fig.10(2). The coordinates of the location of the first red line from top to bottom indicate that, the base sequence of the molecule d from the 1th to the 15th sites at 5'- 3' direction is paired with all of the fifteen bases of the molecule t0s0 at 3'- 5' direction. The coordinates of the location of the second red line from top to bottom indicate that, the base sequence of the molecule d from the 16th to the 33th sites at 5'- 3' direction is paired with all of the eighteen bases of the molecular t0u1 at 3'- 5' direction. The coordinates of the location of the third red line from top to bottom indicate that, the base sequence of the molecule d from the 34th to the 51th sites at 5'- 3'

Tab.2 checking for $\varphi_1$: the designed encoding sequence, where WC means Watson-Crick complementary strand of code

| | |
|---|---|
| code | 5' GCCAGAATTGCAAGGCAGCGAATTGCAAGGCGCGGAATTGCAAGGCCCCGAATTGCAAGGCCGTCCGACGC 3' |
| WC | 3' CGGTCTTAACGTTCCGTCGCTTAACGTTCCGCGCCTTAACGTTCCGGGGCTTAACGTTCCGGCAGGCTGCG 5' |

Tab.3. checking for $\varphi_1$: the encoding rules of input strings charactering runs, encoding by the way of sticker automata

| Object of code | DNA code |
|---|---|
| Initiator sequence | $I_1$ = 5' GCCA 3' |
| Spacer sequence | $X_0$ = 5' GAA 3', $X_1$ = 5' TTG 3', $X_2$ = 5' CAA 3', $X_3$ = 5' GGC 3' |
| Terminator sequence | $I_2$ = 5' CGTC 3' |
| Atomic proposition | p= 5' CGA 3', q=5' CCC 3', r=¬p=5' CGC 3', s=¬q=5' AGC 3', u=r∧s=5' GCG 3' |

Tab.4. checking for $\varphi_1$: the encoding of FSA $A_1$ of formula, encoding by the way of sticker automata, where strikeouts mean WC

| Object of code | Abbreviated transition rule | DNA code |
|---|---|---|
| Initial state $s_0$ | none | 3' ~~I1X0~~ 5' = 3' CGGTCTT 5' |
| Acceptance state $s_2$ | none | 3' ~~X3I2~~ 5' = 3' CCGGCAG 5' |
| Transition rule $t(s_0,s)=s_0$ | t0s0 | 3' ~~X1X2X3 s X0~~ 5' = 3' AACGTTCCGTCGCTT 5' |
| Transition rule $t(s_0,u)=s_1$ | t0u1 | 3' ~~X1X2X3 u X0X1~~ 5' = 3' AACGTTCCGCGCCTTAAC 5' |
| Transition rule $t(s_0,s)=s_2$ | t0s2 | 3' ~~X1X2X3 s X0X1X2~~ 5' = 3' AACGTTCCGTCGCTTAACGTT 5' |
| Transition rule $t(s_1,s)=s_1$ | t1s1 | 3' ~~X2X3 s X0X1~~ 5' = 3' GTTCCGTCGCTTAAC 5' |
| Transition rule $t(s_1,q)=s_2$ | t1q2 | 3' ~~X2X3 q X0X1X2~~ 5' = 3' GTTCCGGGGCTTAACGTT 5' |

Tab.5 the runs of the system $M_1$

| path | DNA code of the path or sequence of nodes (atomic propositions) crossed by the path |
|---|---|
| Code of path 1 | GCCA GAATTGCAAGGC AGC GAATTGCAAGGC AGC\|GCG GAATTGCAAGGC CCC GAATTGCAAGGC CGTC |
| sequence of nodes crossed by path 1 | 0,1,2 (s,s\|u,q) |
| Code of path k | GCCA GAATTGCAAGGC (AGC GAATTGCAAGGC AGC\|GCG GAATTGCAAGGC)$^k$ CCC GAATTGCAAGGC CGTC |
| sequence of nodes crossed by path k | $(0,1)^k,2$ |

direction is paired with all of the eighteen bases of the molecular t1q2 at 3'- 5' direction. This phenomenon suggests that, the complete double stranded DNA molecules are formed, and the hybridization among the four kinds of single stranded DNA molecules is specific.

Comparing the color of the three red lines with the color change of the vertical bar on the right side of Fig.10(2), we can see clearly that the former colors are very close to the color at the top of the vertical bar. This phenomenon suggests that the probabilities of these base pairs are close to 100%. It is a higher degree of specificity.

As shown in Fig.10(1), the concentration of the molecule strand1-strand2-strand3-strand4 is 100uM, and the concentrations of the molecules t0s0, t0u1, t1q2 and d are approximately equal to 0, after their hybridization. It indicates that all of the molecular reactants are involved in the specific hybridization, due to 100uM/100uM=100%. Therefore, the false negative rate is about 0, the false positive rate is approximately equal to 0, the true positive rate is approximately equal to 100%. Once again, it suggests that the four kinds of molecules are hybridized with strong specific.

 (2) Group 2: t0s0, t0u1, t0s2 and d

The concentrations of the four kinds of molecules reach respectively 100uM, 100uM, 100uM and 100uM, and their molecular numbers are respectively 60000, 60000, 60000 and 60000. With the temperature naturally dropped to 10 Celsius degree, the hybridization reaction is observed. Fig.10(3) shows the result of the hybridization, where strand1 means d, strand2 means t0s0, strand3 means t0u1 and strand4 means t0s2.

See Fig.10(3). There exists a red dot in the segment of strand1 of the vertical thin bar on the right side of the strand4, indicating that some bases of strand1 do not be paired. It suggests that the four kinds of molecules do not form the complete double strands.

 (3) Group 3: t0s0, t0u1, t1s1 and d

All of the biochemical conditions and the processes are similar with the ones of the above groups. Fig.10(4) shows the result. There exist some red dots in the segment of strand1 of the vertical thin bar on the right side of the strand4, suggesting that the four kinds of molecules do not form the complete double strands.

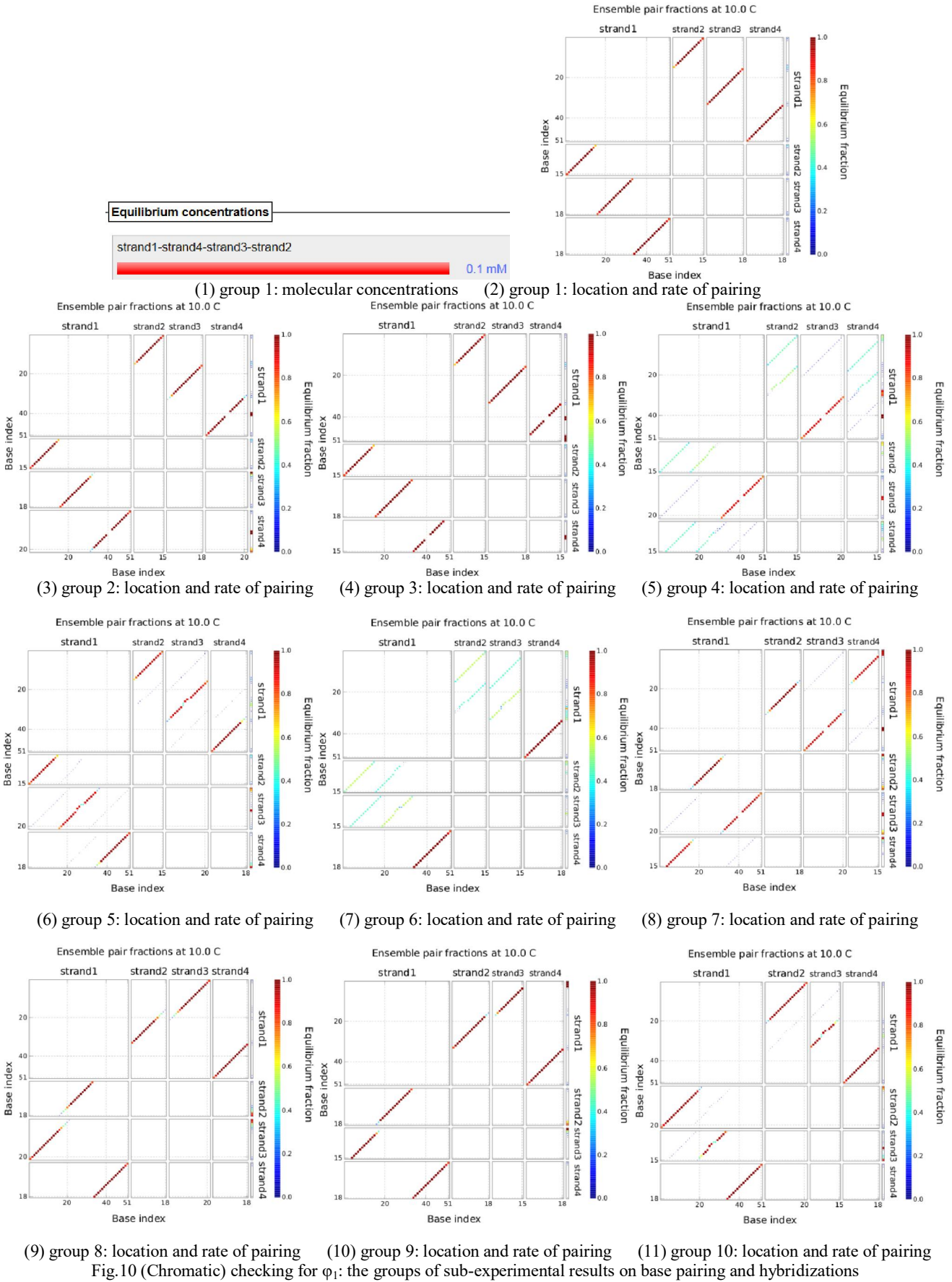

(1) group 1: molecular concentrations    (2) group 1: location and rate of pairing

(3) group 2: location and rate of pairing    (4) group 3: location and rate of pairing    (5) group 4: location and rate of pairing

(6) group 5: location and rate of pairing    (7) group 6: location and rate of pairing    (8) group 7: location and rate of pairing

(9) group 8: location and rate of pairing    (10) group 9: location and rate of pairing    (11) group 10: location and rate of pairing

Fig.10 (Chromatic) checking for $\varphi_1$: the groups of sub-experimental results on base pairing and hybridizations

(4) Group 4: t0s0, t0s2, t1s1 and d

Tab.6 the results: checking for $\varphi_1$ in the different paths of $M_1$ (whether or not the path satisfies $\varphi_1$)

| formula | Path 1 | Path k, where 15>k>1 | Path 15 | Does $M_1$ satisfy $\varphi_1$? |
|---|---|---|---|---|
| $\varphi_1$ | yes | yes | yes | yes |

Tab.7 the model checking results: $M_1$ and the basic CTL constructs (whether or not the system $M_1$ satisfies these formulas)

| formula | Result | The used algorithm and decision basis |
|---|---|---|
| ApUq | No | TL-MC-DNA determine $M_1$ doesn't satisfy pUq, thus algorithm 1 determine $M_1$ doesn't satisfy ApUq |
| AFp | Yes | TL-MC-DNA determine $M_1$ satisfies Fp, thus algorithm 1 determine $M_1$ satisfies AFp |
| AGp | No | TL-MC-DNA determine $M_1$ doesn't satisfy Gp, thus algorithm 1 determine $M_1$ doesn't satisfy AGp |
| AXp | No | TL-MC-DNA determine $M_1$ doesn't satisfy Xp, thus algorithm 1 determine $M_1$ doesn't satisfy AXp |
| EpUq | No | Extended TL-MC-DNA determine $M_1$ satisfies $\varphi_1$, thus algorithm 2 determine $M_1$ doesn't satisfy EpUq |
| EFp | Yes | TL-MC-DNA determine $M_1$ doesn't satisfy G¬p, thus algorithm 2 determine $M_1$ satisfies EFp |
| EGp | No | TL-MC-DNA determine $M_1$ satisfies F¬p, thus algorithm 2 determine $M_1$ doesn't satisfy EGp |
| EXp | No | TL-MC-DNA determine $M_1$ satisfies X¬p, thus algorithm 2 determine $M_1$ doesn't satisfy EXp |

All of the biochemical conditions and the processes are similar with the ones of the above groups. Fig.10(5) shows the result. None of red line is found at the 5' end of strand1, indicating that the 5' end of strand1 does not be paired with any molecule. It suggests that the four kinds of molecules do not form the complete double strands.

(5) Group 5: t0s0, t0s2, t1q2 and d

All of the biochemical conditions and the processes are similar with the ones of the above groups. Fig.10(6) shows the result. There exist some red dots in the segments of strand3 and strand4 of the vertical thin bar on the right side of the strand4, suggesting that the four kinds of molecules do not form the complete double strands.

(6) Group 6: t0s0, t1s1, t1q2 and d

All of the biochemical conditions and the processes are similar with the ones of the above groups. Fig.10(7) shows the result. None of red line is found at the 5' end of strand1, suggesting that the four kinds of molecules do not form the complete double strands.

(7) Group 7: t0u1, t0s2, t1s1 and d

All of the biochemical conditions and the processes are similar with the ones of the above groups. Fig.10(8) shows the result. There exist some red dots in the segments of strand1 of the vertical thin bar on the right side of the strand4, suggesting that the four kinds of molecules do not form the complete double strands.

(8) Group 8: t0u1, t0s2, t1q2 and d

All of the biochemical conditions and the processes are similar with the ones of the above groups. Fig.10(9) shows the result. There exist some red dots in the segments of strand2 and strand3 of the vertical thin bar on the right side of the strand4, suggesting that the four kinds of molecules do not form the complete double strands.

(9) Group 9: t0u1, t1s1, t1q2 and d

All of the biochemical conditions and the processes are similar with the ones of the above groups. Fig.10(10) shows the result. There exist a red dot in the segments of strand1 of the vertical thin bar on the right side of the strand4, suggesting that the four kinds of molecules do not form the complete double strands.

(10) Group 10: t0s2, t1s1, t1q2 and d

All of the biochemical conditions and the processes are similar with the ones of the above groups. Fig.10(11) shows the result. There exist some red dots in the segments of strand2 and strand3 of the vertical thin bar on the right side of the strand4, suggesting that the four kinds of molecules do not form the complete double strands.

According the ten groups of sub-experiments mentioned above, we find that none of group excerpt for the group 1, i.e., t0s0, t0u1, t1q2 and d, can form the complete double strands by the hybridization reaction. That is to say, the systematic run suq satisfies the formula $\varphi_1$, since the first state does not satisfy q, the second state satisfies none of p and q, and the third state satisfies q.

The above results are gotten when k=1. Ref.[20] has proved that a system satisfies the formula pUq, if and only if all the runs whose lengths are less than $|V|*2^{|V|-1}+|E|$ satisfy pUq, where $|V|$ and $|E|$ mean the number of nodes and the number of edges in the systematic FSA, respectively. Similarly we can prove that this conclusion holds for $\varphi_1$. $M_1$ has three nodes and three edges, thus we needs to check fifteen paths due to k=3*$2^{3-1}$+3=15. With the same experimental way, we have checked the $k^{th}$ path, as shown in Tab.6. $M_1$ satisfies the formula $\varphi_1$ since all paths, i.e., runs, satisfy this formula.

Tab.8 A comparison of power among the various DNA model checking methods
(Does the method can conduct DNA model checking for a given formula?)

| Logic | Basic construct | Method in [18] | Method in [20] | Method in [19] | The new method |
|---|---|---|---|---|---|
| LTL | pUq | No | Yes | Yes | No |
| | Fp | No | Yes | The method can be used to check. However, it is not practical to check due to the limitation of the code. | No |
| | Gp | No | Yes | The method can be used to check. However, it is not practical to check due to the limitation of the code. | No |
| | Xp | No | Yes | No | No |
| CTL | ApUq | No | No | No | Yes |
| | AFp | No | No | No | Yes |
| | AGp | No | No | No | Yes |
| | AXp | No | No | No | Yes |
| | EpUq | No | No | No | Yes |
| | EFp | Yes | No | No | Yes |
| | EGp | No | No | No | Yes |
| | EXp | No | No | No | Yes |

By calling the procedure for checking $\varphi_1$, the algorithm 2 can get the model checking results on the formula EpUq. The model checking results on the basic CTL constructs are shown in Tab.7.

According to the experimental processes and results in section 4.1, we can safely say, the algorithm 3, which can be employed to check the basic CTL constructs, has been effectively implemented in molecular biology.

### 4.2 Some comparisons among the new method and the related ones

Tab.8 gives a comparison of power between the new method and the existing ones.

Ref.[18] proposed a DNA-computing-based approach for checking the basic CTL construct EFp. However, this method cannot deal with other basic CTL constructs. In comparison, the new method can conduct model checking for all of the basic CTL constructs via some DNA molecules.

Ref.[19] and Ref.[20] gave the DNA-computing-based approaches for checking all of the basic LTL formulas and some popular LTL formulas. However, this method can deal with none of CTL formula. In comparison, the new method can do it.

# V Conclusions

Early researches on DNA computing focus on the models and algorithms based on non autonomous. In recent years, the DNA computing technique has developed to the self-assembly. The main results of this paper are the algorithm 3, which is based on the self assembly of sticker automata. With these algorithms at hands, we can conduct model checking for the basic CTL constructs via some DNA molecules. Furthermore, Temporal Logic of Actions [23] is another popular temporal logic. And the author of this paper has performed DNA model checking for the basic formulas of TLA, in the relatively similar way with the one presented in this paper. More details are not given here due to the limitation of the space. This is a beneficial for us using the newly proposed method.

# Acknowledgements

The initial idea presented in this paper and the ideas on principles of DNA model checking for the several kinds of temporal logics, came up to my mind in 2010. Hereafter, I gradually realized these ideas one by one, and continually tried to improve these works, especial the full and sound simulated experiments, as well as some effective ways of reading results. In this process, my supervisors, my postgraduate students, and some respected peers, including a famous leading scholar, gave me a number of constructive and meaningful comments and suggestions on these works. I'd like to thank all of them.

This work has been supported by the National Natural Science Foundation of China under Grant U1204608 and 61572444.

The arXiv: 1608.01785v1 was submitted on Aug, 5, 2016, and the arXiv: 1608.01785v3, i.e., this version of this paper, was submitted on Feb, 20, 2017.